\documentclass[prd,preprint,aps,amsfonts,amssymb,nofootinbib,tightenlines]{revtex4}

\usepackage{amscd}
\usepackage{amsmath}
\usepackage{bm}
\usepackage[dvips]{color}                 
\usepackage{graphicx}
\usepackage{eufrak}
\usepackage{eucal}

\newcommand{\beq}{\begin{equation}}
\newcommand{\eeq}{\end{equation}}
\newcommand{\bea}{\begin{eqnarray}}
\newcommand{\eea}{\end{eqnarray}}

\newcommand{\nn}{\nonumber}
\newcommand{\benn}{\begin{displaymath}}
\newcommand{\eenn}{\end{displaymath}}
\newcommand{\ket}[1]{| #1 \rangle}                     

\def\[{\left[}
\def\]{\right]}

\begin{document}

\title{Phase Transition in Imbalanced Fermion Superfluids}

\author{Heron Caldas\footnote{Email: {\tt hcaldas@ufsj.edu.br}.}}
\affiliation{Universidade Federal de S\~{a}o Jo\~{a}o del Rey, S\~{a}o Jo\~{a}o del Rei, 36300-000, MG, Brazil.}

\begin{abstract}
In this chapter the recent theoretical work on phase transition in imbalanced fermion superfluids is reviewed. The imbalanced systems are those in which the two fermionic species candidate to form pairing have different Fermi surfaces or densities. We consider systems subjected to weak interactions. In this scenario two distinct phase transitions are predicted to occur. A thermodynamical phase transition, induced by the temperature (T), and a quantum phase transition as a function of the increasing chemical potentials asymmetry, that takes place at zero temperature. We also briefly discuss some recent experimental work at non-zero T with imbalanced Fermi gases in cold atomic traps.

\end{abstract}
\maketitle
\bigskip

\section{Introduction}

It is well known that fermions with opposite spins near a common Fermi surface suffer the Cooper pairing instability at arbitrarily weak coupling, below a certain critical temperature, leading to the phenomenon of superfluidity. When the two fermion species have the same density, the ground state is described by the successfully Bardeen-Cooper-Shrieffer (BCS)~\cite{BCS} theory of superconductivity. A novelty in the pairing mechanism is brought about by an imbalance in the number densities of the two species\footnote{These different species could be two fermionic atoms ($^{40}{\rm K}$ or $ ^{6}{\rm Li}$) or hyperfine states of the same atom~\cite{Marco,Thomas,Granade}. The nucleus of neutron stars could also have the basic ingredients for asymmetric pairing: quarks with different flavors \cite{krishna_review,Rischke:2003mt,Schafer:2003vz}.}: certainly there will be fermions without a partner. 

For this asymmetrical scenario various exotic phases have been suggested, such as the first proposal\footnote{Sarma considered a model exhibiting asymmetry between the (fixed) chemical potentials of two particle species at zero and finite temperature \cite{Sarma}.} by Sarma~\cite{Sarma}, the Larkin, Ovchinnikov, Fulde and Ferrel (LOFF)-phase~\cite{larkin} (where the gap parameter breaks translational invariance), deformed Fermi surfaces~\cite{Sedrakian1,Sedrakian2} (in which the Fermi-surfaces of the two species are deformed into ellipsoidal form at zero total momentum of Cooper pairs), the breached pair superfluid phase (BP)~\cite{Liu1,Wilczek3} (composed by a homogeneous mixture of normal and superfluid phases), and phase separation in real space~\cite{PRL1,Heron} (defined as an inhomogeneous mixed phase formed by normal and superfluid components). Alternatives beyond mean field have also been presented, such as the (induced) P-wave superfluidity~\cite{Aurel}.

The very recent experimental demonstration of superfluidity through the observation of vortices~\cite{Martin}, and phase separation with controlled population imbalance in atomic gases~\cite{Ketterle,Hulet} have promoted a great interest in the field, including prospects from the theoretical point of view~\cite{Paolo,Mizushima,Dan,Yi,Silva,Hu,Melo,Machida}. In these experiments, the strength of the interactions can be tuned by the use of the Feshbach resonance, varying the external magnetic field. In this way, the crossover from BCS superfluidity to the Bose-Einstein condensation (BEC) can be accessed. For a more complete description on recent experimental work on various novel superfluid phases in fermion systems, see the chapter {\it Realization, characterization, and detection of novel superfluid phases with pairing between unbalanced fermion species} in this volume, Ref.~\cite{Kun}.

In passing we note that the counterpart for ordering in the quantum chromodynamics (QCD) context (for the case where the quarks have approximately the same Fermi surface) at high baryon density, is the color superconductivity \cite{krishna_review,Alford:2001dt,Schafer:2003vz,Rischke:2003mt}. For high density asymmetric quark matter the analog of LOFF\footnote{A review of the theoretical approach and phenomenological applications of the LOFF state in Condensed Matter and QCD, can be found in Ref.~\cite{Nardulli}. An analysis of the competition between the BP and LOFF pairing mechanisms in asymmetric fermion superfluids is shown in Ref.~\cite{He2}.} state leads to crystalline color superconductivity \cite{QCD-2}. In QCD with two light flavors, the ground state is the two-flavor color superconductor (2SC) \cite{Alford}. The pairing between $u$ and $d$ quarks under the condition of charge neutrality and $\beta$-equilibrium at intermediate baryon densities, denoted gapless color superconductivity, has been studied in Refs. \cite{Igor1,Igor2}. The quark (and nuclear) pairing will be discussed in other chapters of this issue.

\section{The Model}

To begin with, let us consider a nonrelativistic dilute (i.e., the particles interact through a short-range attractive interaction) cold fermionic system, described by the following Hamiltonian
\begin{equation}
\label{eq-1}
{\cal H}=H-\sum_{k,\alpha}\mu_{\alpha}n_{\alpha}=\sum_{k} {\epsilon}^{a}_k a^{\dagger}_{k} a_{k}+{\epsilon}^{b}_k b^{\dagger}_{k} b_{k} - g \sum_{k,k'} a^{\dagger}_{k'} b^{\dagger}_{-k'} 
b_{-k} a_{k},\,
\end{equation}
where $a^{\dagger}_{k}$, $a_k$ are the creation and annihilation operators for the $a$ particles (and the same for the $b$ particles)
and ${\epsilon}^{\alpha}_k$ are their dispersion relation, defined by ${\epsilon}^{\alpha}_k=\xi_k^{\alpha}-\mu_{\alpha}$,
with $\xi_k^{\alpha}=\frac{k^2}{2m_{\alpha}}$ and $\mu_{\alpha}$ being the chemical potential of the (non-interacting) $\alpha$-specie, $\alpha=a,b$. To reflect an attractive (s-wave) interaction between particles $a$ and $b$ we take $-g<0$.

\subsubsection{The Thermodynamic Potential}

We now derive the thermodynamic potential at fixed chemical potentials\footnote{The situation where the chemical potentials are kept fixed can find place, for instance, in a gas of fermionic atoms connected to reservoirs of species $a$ and $b$, so the number densities are allowed to change in the system.}, and the finite temperature gap equation for an asymmetrical fermion system, in order to determine the critical temperature. We follow the usual derivation of the textbooks~\cite{Feynman}, however extending the analysis for the imbalanced systems we are investigating. The thermodynamic potential is given by

\begin{equation}
\label{ap2}
\Omega=E-TS,
\end{equation}
where $E$ is the internal energy and $S$ the entropy. Let us define $f_k$ as the probability of an $a$ particle with momentum $\bf{k}$ is excited, and similarly $g_k$ as the probability of a $b$ particle with momentum $\bf{-k}$ is excited. One can write the following possible probabilities for the particles states: The probability that a given pair of $\bf{k}$ is unexcited is
\begin{equation}
\label{p1}
P_k(0)=(1-f_k)(1-g_k).
\end{equation}
The probability that one of the states ($\bf{k}$, for instance) is excited, and the other is not, is
\begin{equation}
\label{p2}
P_k(2)=f_k(1-g_k).
\end{equation}
Now the probability that the state $-\bf{k}$ is excited, and the other is not, is
\begin{equation}
\label{p3}
P_k(3)=(1-f_k)g_k.
\end{equation}
And finally, the probability that both states are excited is
\begin{equation}
\label{p4}
P_k(1)=f_kg_k.
\end{equation}
The entropy is defined as

\begin{equation}
\label{ent}
S=-\sum_{k} {\rm probability~ of~ state~ i} \times \ln [ {\rm probability~ of~ state~ i}]=-\sum_{k} P_i \ln(P_i),
\end{equation}
where we have set the Boltzmann constant equal to one. It is left as an exercise to the reader to show that $S$ for an asymmetrical fermion gas is found to be

\begin{equation}
\label{ap1}
S=-\sum_{k} \left\{ f_k \ln(f_k) + (1-f_k) \ln(1-f_k) + g_k \ln(g_k) + (1-g_k) \ln(1-g_k) \right\}.
\end{equation}
The BCS ground state, which describes a superposition of empty and occupied (paired) states, is given by~\cite{BCS}
\begin{equation}
\label{bcs}
\left| BCS \right\rangle=\prod_{k}\left[u_k+v_k a^{\dagger}_{k} b^{\dagger}_{-k}\right]\left|0\right\rangle ,
\end{equation}
where the arbitrary complex (a priori) coefficients $u_k$ and $v_k$ are to be determined by a variational calculation. They are subjected to the
normalization, $|u_k|^2+|v_k|^2=1$, and spin singlet, $u_k=u_{-k}$, $v_k=v_{-k}$, conditions. At zero temperature, the internal or ground state energy $E$ is simply the expectation value of the Hamiltonian, $E \equiv <{\rm BCS}|{\cal H}|{\rm BCS}>$. At finite temperature the energy has to take into account the excitations probabilities, $E=\sum_{k} E_i P_i$~\cite{Feynman}. Then we find

\bea
\label{ap3}
E=\sum_{k} \left\{ \epsilon_k^a [ (1- f_k -g_k)u_k^2+ f_k] + \epsilon_k^b [ (1- f_k -g_k)u_k^2+ g_k] \right\}\\
\nonumber 
-g \sum_{k,k'} u_{k'} v_{k'} u_k v_k (1-f_k - g_k)(1-f_{k'} - g_{k'}).
\eea
Plugging Eqs.~(\ref{ap1}) and (\ref{ap3}) in Eq.~(\ref{ap2}), and taking the minimizations

\bea
\label{ap4}
\frac{\delta \Omega}{\delta f_k}=0,\\
\nonumber
\frac{\delta \Omega}{\delta g_k}=0,\\
\nonumber
\frac{\delta \Omega}{\delta u_k}=0,\\
\nonumber
\eea
we find, respectively,

\begin{equation}
\label{ap5}
f_k=1/(e^{\beta {\cal{E}}_k^a}+1),
\end{equation}

\begin{equation}
\label{ap6}
g_k=1/(e^{\beta {\cal{E}}_k^b}+1),
\end{equation}

\begin{equation}
\label{ap7}
u_k^2=\frac{1}{2} \left(1+ \frac{\epsilon_k^{+}}{E_k} \right),
\end{equation}
where 
\begin{equation}
\label{qpe}
{\cal{E}}_k^{a,b}=\pm \epsilon_k^{-} + E_k,
\end{equation}
are the quasiparticle excitations, with $E_k=\sqrt{ {\epsilon_k^{+}}^2+\Delta^2(T) }$ and $\epsilon_k^{\pm} \equiv \frac {\epsilon_k^a \pm \epsilon_k^b}{2}$. In terms of $f_k$, $g_k$ and $u_k$ the thermodynamic potential becomes

\begin{equation}
\label{tp}
\Omega=\frac{\Delta^2}{g} + \sum_{k} \left[\epsilon_k^{+}-E_k-T \ln(e^{-\beta {\cal{E}}_k^{a}}+1) -T \ln(e^{-\beta {\cal{E}}_k^{b}}+1) \right].
\end{equation}
In the definition of  ${\cal{E}}_k^{a,b}$ we have also defined

\begin{equation}
\label{ap8}
\Delta(T)=g \sum_{k} u_k v_k (1-f_k - g_k).
\end{equation}
Since $v_k^2=1-u_k^2$, then $u_k v_k=\frac{\Delta}{2 E_k}$, and the gap equation can be written as

\begin{equation}
\label{ap9}
1=g \sum_{k} \frac{1}{2 E_k} \left( 1-f_k - g_k \right).
\end{equation}
We note that the equation above can also be obtained by $\frac{\partial \Omega}{\partial \Delta}=0$. 

\section{Phase Transitions}

\subsection{Phase transitions in fully gapped systems}
 
\begin{figure}[t]
\includegraphics[height=3in]{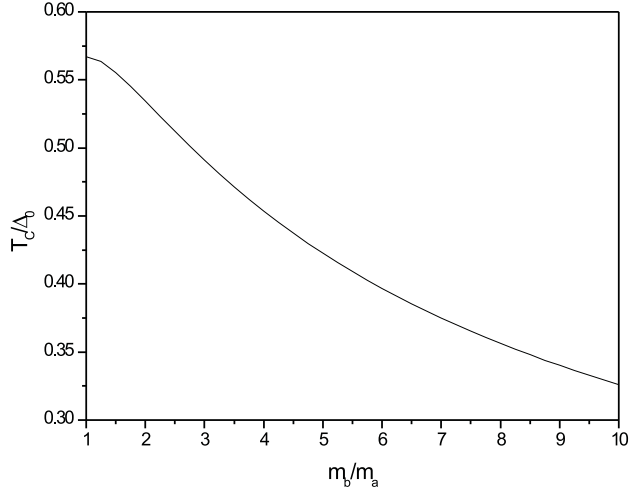}
\caption{\label{TempC1}\textit{$T_c / \Delta_0$ as a function of $m_b/m_a$ for a system constrained to $m_a \mu_a =m_b \mu_b$. } }
\end{figure}

The critical temperature $T_c$ is, by definition, the temperature at which $\Delta=0$. Then Eq.~(\ref{ap9}) becomes

\begin{equation}
\label{ap10}
1=g \sum_{k} \frac{1}{\epsilon_k^a+\epsilon_k^b} \left(1-\frac{1}{e^{\beta_c \epsilon_k^a}} - \frac{1}{e^{\beta_c \epsilon_k^b}} \right).
\end{equation}
After the momentum integration, Eq.~(\ref{ap10}) can be written as~\cite{Heron3}

\beq
\label{tc1}
\frac{1}{g \rho (0)} - \ln \left(\beta_c \sigma \frac{\omega_D}{\pi} \right)=-\frac{1}{2} {\cal F} (a),
\eeq
where $\rho(0)=\frac{M k_F}{ \pi^2}$ is the density of states at the Fermi level, with $k_F=\sqrt{2 M \mu}$ being the ``average'' Fermi surface having $\mu \equiv \mu_a + \mu_b$, and $M=\frac{m_a m_b}{m_a + m_b}$ is the reduced mass. We also introduced $\sigma \equiv \frac{M}{\sqrt{m_a m_b}}$ as a dimensionless parameter reflecting the mass or chemical potential asymmetry, and ${\cal F} (a)= \Psi (\frac{1}{2}+\frac{ia}{\pi })+\Psi (\frac{1}{2}-\frac{ia}{\pi })$, with $\Psi$ being the digamma function, defined as $\Psi(z)=\frac{\Gamma'(z)}{\Gamma(z)}$, where z is a complex number with a positive real component, $\Gamma$ is the gamma function, and $\Gamma'$ is the derivative of the gamma function and $a \equiv \frac{\beta}{2} \eta = \frac{\beta}{2} \frac{m_b \mu_b- m_a \mu_a}{m_a + m_b}$. Eq.~(\ref{tc1}) can be put in the form

\beq
\label{tc2}
T_c=\frac{\sigma\Delta_0}{2 \pi} e^{-\frac{1}{2} {\cal F}(a_c)},
\eeq
where $a_c=\frac{\beta_c}{2} \eta $ and $\Delta_0=2\omega_D~ e^{-1/\rho(0)g}$ is the BCS gap parameter in the weak coupling limit, $\rho(0)g<<1$. The critical temperature we have obtained refers to the situation where all the fermions are gapped\footnote{By construction the $a$ and $b$ particles in the BCS state have the same number densities, see Eq.~(\ref{bcs}).}; they have the same Fermi surface in spite of the masses and chemical potentials asymmetry. The consequence is that $P^a_F=P^b_F$ (where $P_F^{a,b}=\sqrt{2 m_{a,b} \mu_{a,b}}$) implies $a_c=0$, resulting in

\beq
\label{tc4}
T_c(P^a_F=P^b_F)= 2\sigma \frac{e^{\gamma}}{\pi} \Delta_0,
\eeq
where we have used that ${\cal F}(0)=-2\gamma -4\ln (2)$, where $\gamma$ is the Euler's constant. An important feature of Eq.~(\ref{tc4}) is that, although Eq.~(\ref{ap10}) requires regularization, the regulator dependence cancels from the result~(\ref{tc2})~\cite{Heron3}.  In the symmetric limit, namely $m_a=m_b$ (or $\mu_a = \mu_b$) the standard BCS relation $T_c/\Delta_0=\frac{e^{\gamma}}{\pi}$~\cite{BCS,Tinkham} is recovered. We can observe from Eq.~(\ref{tc4}) that the critical temperature for the system constrained to $P^a_F=P^b_F$ (or $m_a \mu_a =m_b \mu_b$) goes with $2 \sqrt{\frac{m_a}{m_b}}~\frac{e^{\gamma}}{\pi}\Delta_0$ for $m_b$ greater than $m_a$ and approaches zero for $m_b>>m_a$. This show that the pair formation is disfavored for very large mass asymmetry, even in systems were the Fermi surfaces match. In Fig.~(\ref{TempC1}) we show the ratio $T_c / \Delta_0$ as a function of $m_b/m_a$. As one can see, $T_c / \Delta_0$ is a smooth function of the mass asymmetry, and goes to zero for $m_b/m_a \to \infty$.

\subsection{Phase transitions in gapless systems}

\begin{figure}[t]
\includegraphics[height=2in]{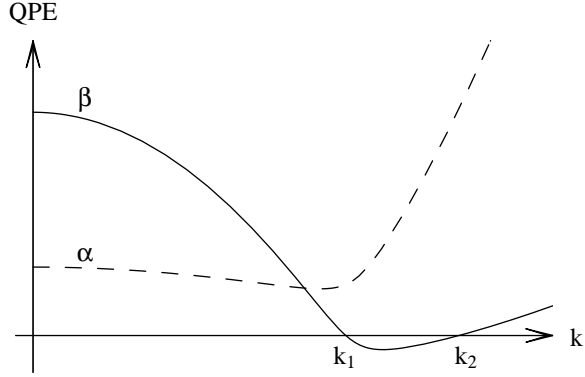}
\caption{\label{fig1}\textit{  Dispersion relation for the quasi-particles
$\alpha$ and $\beta$ showing a region where ${\cal E}^\beta_k$ is negative for
$m_b= 7 m_a$, $P_F^b=1.45\, P_F^a$, and $\Delta=0.29\Delta_0$, obtained by Eq. (\ref{eq26}). Solid curve corresponds to ${\cal E}_k^\beta$
and dashed curve corresponds to ${\cal E}_k^\alpha$.} }
\end{figure}

Depending on the relative magnitudes of the particles Fermi surfaces and masses, the quasiparticle excitations ${\cal{E}}_k^{a,b}$ can be negative. If we choose $P^b_F > P^a_F$ and $m_b \geq m_a$, only ${\cal{E}}_k^{b}$ will cross zero. From the equation $\epsilon_k^a \epsilon_k^b = - \Delta^2$ we determine the roots of ${\cal{E}}_k^{b}$~\cite{Wu,Wilczek2,PRL1,Heron}: 

\begin{equation}
\label{roots}
k_{1,2}^2= \frac{1}{2} {\delta P_F^+}^2 \mp \frac{1}{2} [{(\delta P_F^-}^2)^2 - 16 m_a m_b \Delta^2]^{1/2},
\end{equation}
where ${{\delta P_F}^{\pm} }^2 \equiv {P_F^b}^2 \pm {P_F^a}^2$. The negativity of ${\cal{E}}_k^{b}$ between $k_1$ and $k_2$ means that the corresponding states (the $b$ particles) are singly occupied. In Fig.~(\ref{fig1}) we show the quasiparticle excitations (QPE) as a function of the momentum $k$. For some values of $\Delta$, ${\cal E}^\beta_k$ may be negative for momenta $k_1\le k\le k_2$.

We can observe from Eq.~(\ref{roots}) that $\Delta$ has a critical value

\begin{equation}
\label{DeltaC}
\Delta_c=\frac{|{\delta P_F^-}^2|}{4 \sqrt{m_a m_b}}.
\end{equation}
For $\Delta > \Delta_c$, $k_{1,2}$ are not real and ${\cal{E}}_k^{a,b}$ never crosses zero. This corresponds to the standard BCS with pairing for all $k$. The situation where $\Delta < \Delta_c$, named ``Sarma phase'', was first pointed out in Ref.~\cite{Sarma}.

The thermodynamic potential in the Sarma phase is obtained when we find a state $\left| \Psi\right\rangle$ which minimizes the internal energy. The smallest energy is reached when the modes with negative ${\cal E}^{a,b}_k$ are filled and the remaining modes are left empty. More precisely, the ground state $\ket{\Psi}$ satisfies~\cite{PRL1,Heron}
\begin{align}
a_k,b_k \ket{\Psi}  &= 0 \quad \text{if}
\quad {\cal E}^{a}_k >0,\nn\\
a_k^\dagger,b_k^\dagger \ket{\Psi} &= 0 \quad \text{if}
\quad {\cal E}^{b}_k <0.
\end{align}
This state can be written in terms of the $a^{\dagger}_{k}$ and $b^{\dagger}_{k}$ operators and the vacuum state $\ket{0}$ as
\begin{equation}
\label{statesarma}
\left| \Psi \right\rangle=\prod_{\substack{k<k_1 \\ k>k_2}}\left[u_k+v_k a^{\dagger}_{k} b^{\dagger}_{-k}\right] \prod_{k_1}^{k_2}b^{\dagger}_{k}\left|0\right\rangle .
\end{equation}
The state above corresponds to having BCS pairing in the modes $k$ where ${\cal E}_k^{a}>0$ and a state filled with particles $b$ in the modes where ${\cal E}_k^{b}<0$. Using this state in the computation of the entropy and the internal energy, the thermodynamic potential of the Sarma phase turns out to be

\bea
\label{SarmaE}
\Omega(g, T, \Delta < \Delta_c) &\equiv& \Omega^{S}(T)\\
\nonumber
&=& \frac{\Delta^2}{g} + \sum_{\substack{k<k_1 \\ k>k_2}} \left[\epsilon_k^{+}-E_k-T \ln(e^{-\beta {\cal{E}}_k^{a}}+1) -T \ln(e^{-\beta {\cal{E}}_k^{b}}+1) \right]\\
\nonumber
&+& \sum_{k_1}^{k_2} \left[\epsilon_k^{b}-T \ln(e^{+\beta \epsilon_k^{b}}+1) \right].
\eea

Since there are gapless states in Eq.~(\ref{SarmaE}) one can not define a critical temperature as performed in the previous situation, i.e., in the fully gapped system. To find the critical temperature ($T_c^*$) in this case, it is necessary a comparison between $\Omega^{S}(T)$ and the normal free energy,  $\Omega^{S}(\Delta=0,T)$, at a given and fixed asymmetry $\delta \mu < \delta \mu_c$\footnote{The prediction for the break down of the fermionic superfluidity is $ \delta \mu_c \equiv \frac{\mu_b-\mu_a}{2}=\Delta_0$~\cite{PRL1,Heron,Carlson}. This picture has been confirmed qualitatively experimentally~\cite{Ketterle}.}, for increasing temperature~\cite{Heron4}. We then define $T_c^*$ for gapless systems as the temperature at which $\Omega(\delta \mu,T=T_c^*,\Delta=\Delta_0)=\Omega(\delta \mu,T=T_c^*,\Delta=0)$. Solving this equality for $T_c^*$ we obtain the transition temperature. 

We point out here that the effect of the temperature in the thermodynamic potential of Eq.~(\ref{SarmaE}) is to induce a transmutation from a stable phase at $\Omega(\delta \mu,T<T_c^*,\Delta=\Delta_0)$ to an also stable phase $\Omega(\delta \mu,T>T_c^*,\Delta=0)$~\cite{Liao,Heron4}. The nature of this transition, however, needs further investigation and will be presented elsewhere~\cite{Heron4}.

\subsection{The quantum phase transition}

The zero temperature limit of Eq.~(\ref{SarmaE}) yields

\bea
\label{SarmaT=0}
\Omega(g,T=0, \Delta < \Delta_c) \equiv \Omega^{S}(T=0) =\frac{\Delta^2}{g} + \sum_{\substack{k<k_1 \\ k>k_2}} \left[\epsilon_k^{+}-E_k\right]+\sum_{k_1}^{k_2} \epsilon_k^{b}.
\eea
With $\frac{\partial \Omega^{S}(T=0)}{\partial \Delta}=0$ (remembering that the partial derivative also hits $k_1$ and $k_2$ in the limits of the integrals) we obtain the gap equation
\begin{equation}
\label{eq22}
1=\frac{g}{2} \int_{\substack{k<k_1 \\ k>k_2}} \frac{d^3 k}{(2 \pi)^3} \frac{1}{\sqrt{{\varepsilon_k^+}^2+\Delta^2}}.
\end{equation}
We can find the gap in the Sarma phase through the identity~\cite{Sarma,Wilczek2,PRL1,Heron}
\begin{equation}
\label{eq23}
\frac{M}{2 \pi |a|}= \int \frac{d^3 k}{(2 \pi)^3} \frac{1}{\sqrt{{\varepsilon_k^+}^2+\Delta_0^2}}=\int_{\substack{k<k_1 \\ k>k_2}} \frac{d^3 k}{(2 \pi)^3}
\frac{1}{\sqrt{{\varepsilon_k^+}^2+\Delta^2}} .
\end{equation}
For small values of the gaps ($\Delta_0, \Delta << \mu_a, \mu_b$) the integrals can be approximated and it is found that
\begin{equation}
\label{eq24}
\frac{\Delta^2}{\Delta_0^2}=\frac{\varepsilon_k^a(k_1)}{\varepsilon_k^a(k_2)},
\end{equation}
which has the solution
\begin{equation}
\label{eq26}
\Delta_S \simeq \sqrt{\Delta_0 \left( \frac{|{\delta P_F^-}^2|}{2 \sqrt{m_a m_b}}-\Delta_0 \right)}.
\end{equation}
It is ease to verify that the Fermi surfaces asymmetry and the gap in the Sarma phase are restricted to:
\begin{eqnarray}
\label{eq27}
2 \sqrt{m_a m_b}\Delta_0& \leq |{\delta P_F^-}^2| \leq &4 \sqrt{m_a m_b}\Delta_0\\
\nonumber
0&\leq\Delta_S\leq&\Delta_0.
\end{eqnarray}

\begin{figure}[t]
\includegraphics[height=2.in]{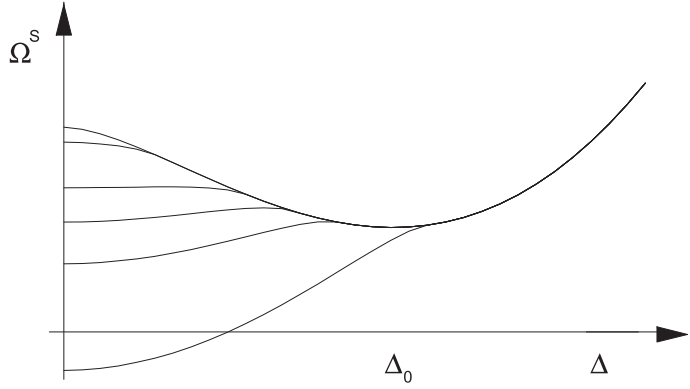}
\caption{\label{omega}\textit{Thermodynamic potential for different values of
$P_F^b$ and $P_F^a$ (constant $k_F$). The top curve corresponds to $P_F^a=P_F^b$ and
the lower curves correspond to increasing values of $|P_F^b-P_F^a|$.} }
\end{figure}

In Fig.~(\ref{omega}) we show the thermodynamic potential as a function of $\Delta$ for different values of $P_F^b$ and $P_F^a$,  keeping the combination $k_F^2/M={P_F^a}^2/m_a + {P_F^b}^2/m_b$ fixed, computed from a numerical evaluation of Eq.~(\ref{SarmaT=0}). As we can see, there exist a special combination of $P_F^a$ and $P_F^b$ for which $\Omega^S(T=0)$ has double minima. Then, given a $P_F^a$, we want to know the correspondent value of $P_F^b$ that satisfies this requirement. We note from Eq.~(\ref{roots}) that when $\Delta=0,~k_1=P_F^a$ and $k_2=P_F^b$, whereas for $\Delta=\Delta_0,~k_1=k_2=\sqrt{\frac{{P_F^a}^2+{P_F^b}^2}{2}}$. Thus, the condition to find the relation between $P_F^a$ and $P_F^b$ is
\begin{align}
\label{min1}
& \Omega^S(T=0,\Delta=0)=\Omega^S(T=0,\Delta=\Delta_0) \to \\
\nonumber
& \int_{\substack{k<P_F^a \\ k>P_F^b}} \frac{d^3k}{(2 \pi)^3} ({\varepsilon}_k^+ - |{\varepsilon}_k^+|)+
\int_{P_F^a}^{P_F^b} \frac{d^3k}{(2 \pi)^3} {\varepsilon}_k^b  =
\int \frac{d^3k}{(2 \pi)^3} ({\varepsilon}_k^+ - E)-\frac{\Delta_0^2}{g}.\
\end{align}
Note that the l.h.s. is just the free-energy of the unpaired $a$ and $b$ particles and the r.h.s. is the standard BCS free energy. Solving both sides we get
\begin{eqnarray}
\label{min2}
\frac{{P_F^a}^5}{m_a}+\frac{{P_F^b}^5}{m_b}=\frac{k_F^5}{ M}+ 15 M k_F  \Delta_0^2,
\end{eqnarray}
where $k_F=[M({P_F^a}^2/m_a + {P_F^b}^2/m_b)]^{1/2}$, and $\Delta_0= \frac{4 \mu}{e^2} e^{\frac{-\pi}{2 k_F |a|}}$ is the standard BCS gap parameter~\cite{Heron}. The solution of the equation above we define as $P_{F, DM}^b$, where $DM$ stands for double minima. Given these fixed values of $P_F^b$ and $P_F^a$ (or $\mu_b$ and $\mu_a$) found above, the system could either be found in the normal, BCS or in a mixed phase, since those states would have the same minimal energy. When the free-energy ${\rm F}(P_{F, DM}^b)$ has its minimum at $\Delta_0$,  after an increase in $P_F^b$ the gap jumps from $\Delta_0$ to $0$, characterizing a first order {\it quantum phase transition} from the superfluid to the normal phase, as has been found theoretically in several physical situations. See, for instance, Refs.~\cite{Norman1,Sedrakian,Paulo}. Summarizing, we find that at fixed chemical potentials the Sarma gap corresponds to an {\it extremum} of the free-energy, it never represents its minimum.

For any curve having momenta $P_F^b \geq P_F^a$ (and fixed in each curve of Fig.~(\ref{omega})) until $P_{F, DM}^b$, both particle species have the same density $n_a=n_b=-\frac{\partial \Omega^S(T=0)}{\partial \mu_{a,b}}=\frac{k_F^3}{6 \pi^2}+\frac{M^2 \Delta_0^2}{2 \pi^2 k_F} \left(5 + \frac{\pi}{ |a| k_F} \right)$. For $P_F^b>P_{F, DM}^b$, there is absorption of particles from the reservoirs so the asymmetry in the number densities gets its maximum value, and the system goes to the normal state with number densities $n_a=\frac{{P_F^a}^3}{6 \pi^2}$ and $n_b=\frac{{P_F^b}^3}{6 \pi^2}$.

The mixed phase (MP) or heterogeneous composition is formed by a normal and a superfluid components, and (differently from the two homogeneous ground states found above) accommodates $n_a$ and $n_b$ particle densities in a trap~\cite{PRL1,Heron}. The MP has been found to be stable, with no surface energy cost at weak coupling~\cite{Heron5}. The issue of the phase transitions in the MP is under investigation~\cite{Heron6}.

\section{Brief Review on Some Aspects of Recent Experimental Work on Fermi Gases at Finite T in Atomic Traps}

For unitary Fermi gases, current experiments~\cite{Kinast1,Kinast2,Kinast3,Kinast4} produce temperatures down to about $0.05T_{F}$, where $T_{F}$ is what the Fermi temperature would be for a noninteracting gas with the same number of atoms and in the same trap conditions. Typically $T_F$ is of order $\mu K$. However, a weakly interacting Fermi gas requires much lower $T$ to achieve superfluidity. For the conditions of these experiments, the mean field approximation with an interaction energy proportional to the scattering length is not valid. However, the mean field approximation with a unitary limit appears approximately valid, furnishing a good agreement with predictions of the collective frequencies, and a very good agreement on the transition temperature~\cite{Thomas2}. 

Recently, measurements of the T-dependent momentum distribution of a trapped Fermi gas consisting of an equal mixture of the two lowest spin states of $^{40}{\rm K}$ in the BCS-BEC crossover regime have been presented~\cite{Chen}. The results show the existence of a competition between the T dependence of the fermionic excitation gap and thermal broadening, leading to non-monotonic behavior in the T dependence of the momentum profiles. Semi-quantitative agreement between theory and experiment using a simple mean-field theory has been found. 

Thus, even when the measurements are done in strongly interacting Fermi gases, mean field theory had qualitatively explained the behavior of these systems, and we expect that our weak coupling mean field BCS results should also be valid, at least, qualitatively.

More recently, direct normal-to-superfluid phase transition has been observed in a strongly interacting Fermi gas with unequal mixtures of the two spin components~\cite{Martin2}. Both the thermodynamical and quantum phase transitions have been detected. A quantum phase transition at $\delta \approx 70 \%$ has been observed, where $\delta =\frac{N_{\uparrow}-N_{\downarrow}}{N_{\uparrow}+N_{\downarrow}}$ is the imbalanced parameter, with $N_{\uparrow}$($N_{\downarrow}$) as the majority(minority) atom number. The authors of Ref.~\cite{Martin2} found that the critical temperature will in general depend on the population imbalance. This is a (expected) guide for theoretical investigations~\cite{Heron6}.

\section{Summary}
In this chapter we have discussed phase transitions in cold fermionic gases composed by two particle species whose Fermi surfaces or densities do not match. We have seen that depending on the relative difference between the Fermi surfaces, two distinct phase transitions are predicted to occur. 

Fully gapped asymmetrical systems undergo a second order phase transition, driven by the temperature, obeying $\frac{\Delta_0}{T_c}=\frac{\pi}{e^{\gamma}} \frac{1}{2\sigma} \approx 1.76 \frac{1}{2\sigma}$, where $\sigma = \frac{\sqrt{m_a m_b}}{m_a + m_b}=\frac{\sqrt{\mu_a \mu_b}}{\mu_a + \mu_b}$ appears to be an {\it universal constant}~\cite{He}. Since experiments are being set up to study pairing between fermions of unequal mass (for example $^{40}{\rm K}$ and $ ^{6}{\rm Li}$), we expect that this expression can be verified soon~\cite{Martin3}.

A system with different Fermi surfaces is found to be in the BCS state while the asymmetry between the chemical potentials is smaller than the critical difference $\delta \mu_c$. A quantum first order phase transition from superfluid to normal phase, at zero temperature, happens as a function of the increasing chemical potentials asymmetry.

As systematic studies at non-zero temperature in imbalanced ultracold systems are just starting, we hope that the work presented in this chapter could stimulate new investigations in this interesting field.

\label{conc}

\section*{Acknowledgments}
The author thanks the Nuclear Science Division of LBL for hospitality and support, and the Institute for Nuclear Theory at the University of Washington for its hospitality, where part of this work was done. He also thanks Paulo Bedaque for enlightening discussions and Martin Zwierlein for helpful comments and for a critical reading of the manuscript. This work was partially supported by the Brazilian agencies CNPq and CAPES.




\begin{thebibliography}{99}

\bibitem{BCS} J. Bardeen, L.N. Cooper, and J. R. Schrieffer, Phys. Rev. {\bf 108}, 1175 (1957).

\bibitem{Marco} B. DeMarco and D.S. Jin, Science {\bf 285}, 1703 (1999).

\bibitem{Granade} S.R. Granade {\it et al.}, Phys. Rev. Lett. {\bf 88}, 120405 (2002).

\bibitem{Thomas} K.~M.~O`Hara {\it et al.}, Science {\bf 298}, 2179 (2002).

\bibitem{krishna_review}
K.~Rajagopal and F.~Wilczek,
\newblock hep-ph/0011333.

\bibitem{Schafer:2003vz}
T.~Schafer,
\newblock hep-ph/0304281.

\bibitem{Rischke:2003mt}
D.~H. Rischke,
Prog. Part. Nucl. Phys. {\bf 52}, 197 (2004).

\bibitem{Sarma} G. Sarma, Phys. Chem. Solid {\bf 24}, 1029 (1963).

\bibitem{larkin} A.I. Larkin and Yu. N. Ovchinnikov, Sov. Phys. JETP {\bf 20} 762 (1965);
P. Fulde and R.A. Ferrel,
\newblock Phys. Rev. {\bf 135}, A550 (1964).

\bibitem{Sedrakian1}  H. Muther, A. Sedrakian, Phys. Rev. Lett. {\bf 88}, 252503 (2002).

\bibitem{Sedrakian2}  A. Sedrakian, J. Mur-Petit, A. Polls, and H. Muther, Phys. Rev. A {\bf72}  013613 (2005).

\bibitem{Liu1} W.~V. Liu and Frank Wilczek, Phys. Rev. Lett. {\bf 90}, 047002 (2003).

\bibitem{Wilczek3}
M.~M. Forbes, E. Gubankova, W.~V. Liu, and F.~Wilczek, Phys. Rev. Lett. {\bf 94} 017001  (2005).

\bibitem{Wu} S.-T. Wu and S.~Yip,
\newblock Phys. Rev. {\bf A67}, 053603 (2003).


\bibitem{PRL1} P. Bedaque, H. Caldas, and G. Rupak, Phys. Rev. Lett. {\bf 91} 247002 (2003).

\bibitem{Heron}
H.~Caldas, Phys. Rev. A {\bf 69} 063602 (2004).

\bibitem{Aurel} A. Bulgac, M.~M. Forbes, and A. Schwenk, cond-mat/0602274.

\bibitem{Martin} M.~W. Zwierlein {\it et al.}, Nature 435, {\bf 1047} (2005).

\bibitem{Ketterle}
M.~W. Zwierlein, A. Schirotzek, C.~H.~Schunck, and W. Ketterle, Science {\bf 311}, 492 (2006).

\bibitem{Hulet}
G.~B.~Partridge {\it et al.}, Science {\bf 311}, 503 (2006).

\bibitem{Paolo} P.~Castorina {\it et al.} Phys. Rev. A {72} 025601 (2005).

\bibitem{Mizushima} T. Mizushima, K. Machida and M. Ichioka, Phys. Rev. Lett {\bf 94}, 060404 (2005). 

\bibitem{Dan} D.~E.~Sheehy and L.~Radzihovsky, Phys. Rev. Lett. {\bf 96}, 060401 (2006).

\bibitem{Yi} W. Yi and L.-M. Duan, cond-mat/0601006.

\bibitem{Silva}  T.~N.~De Silva and E.~J.~Mueller, cond-mat/0601314.

\bibitem{Hu} Hui Hu and Xia-Ji Liu, cond-mat/0603332.

\bibitem{Melo} M. Iskin, C. A. R. Sa de Melo, cond-mat/0604184.

\bibitem{Machida} K. Machida, T. Mizushima, and M. Ichioka, cond-mat/0604339.


\bibitem{Kun} Kun Yang, cond-mat/0603190.


\bibitem{Alford:2001dt}
M.~G. Alford,
\newblock Ann. Rev. Nucl. Part. Sci. {\bf 51}, 131 (2001).

\bibitem{Nardulli} R.~Casalbuoni and G.~Nardulli, Rev. Mod. Phys. {\bf 76}, 263 (2004).

\bibitem{He2}  L.~He, M.~Jin, and P.~Zhuang, cond-mat/0601147,

\bibitem{QCD-2} J. Bowers and K. Rajagopal, hep-ph/0204079.

\bibitem{Alford} M. Alford, K. Rajagopal, and F. Wilkzek, Phys. Lett. B {\bf 422}, 247 (1998).

\bibitem{Igor1} I. Shovkovy and M. Huang, Phys. Lett. B {\bf 564}, 205 (2003).

\bibitem{Igor2} M. Huang and I. Shovkovy, Nucl. Phys. A {\bf 729}, 835 (2003).

\bibitem{Feynman} R.~P.~Feynman, {\it Statistical Mechanics} (Addison Wesley, 1994).

\bibitem{Heron3} H. Caldas, C.~W. Morais, and A.~L. Mota, Phys. Rev. D {\bf72} 045008 (2005). 

\bibitem{Wilczek2}
E.~Gubankova, W. Vincent Liu, and F.~Wilczek, Phys. Rev. Lett. {\bf 91} 032001 (2003).

\bibitem{Tinkham} M. Tinkham, {\it Introduction to Superconductivity} (MacGraw-Hill, New York 1996).

\bibitem{Heron4} H. Caldas {\it et al.}, work in progress.


\bibitem{Norman1}
N.~K. Glendenning,
\newblock Phys. Rev. {\bf D46}, 1274 (1992).

\bibitem{Sedrakian} A. Sedrakian, Phys. Rev. {\bf C 63}, 025801 (2001).

\bibitem{Heron5} H. Caldas, cond-mat/0601148.

\bibitem{Heron6} H. Caldas, work in progress.

\bibitem{Paulo} P. Bedaque, Nuc. Phys. {\bf A 697}, 569 (2002).

\bibitem{Kinast1}  J. Kinast {\it et al.}, Phys. Rev, Lett. {\bf 92}, 150402 (2004).

\bibitem{Kinast2}  J. Kinast, A. Turlapov, and J. E. Thomas, Phys. Rev. A {\bf 70}, 051401 (2004).

\bibitem{Kinast3}  J. Kinast {\it et al.}, Science {\bf 307}, 1296 (2005).

\bibitem{Kinast4}  J. Kinast, A. Turlapov, and J.E. Thomas, cond-mat/0502507.

\bibitem{Thomas2}  J. E. Thomas, private communication.

\bibitem{Chen} Q.~Chen {\it et al.}, cond-mat/0604469.

\bibitem{Martin2}  M.~W.~Zwierlein, C.~H.~Schunck, A.~Schirotzek, and W.~Ketterle, cond-mat/0605258.

\bibitem{Liao} J.~F.~Liao and P.~Zhuang, Phys. Rev. D {\bf 68}, 114016 (2003).

\bibitem{He} L.~He, M.~Jin, P.~Zhuang, cond-mat/0603683.

\bibitem{Martin3} M.~W.~Zwierlein, private communication.

\bibitem{Carlson} J.~Carlson and S.~Reddy, Phys. Rev. Lett. {\bf 95}, 060401 (2005).







\end{thebibliography}
\end{document}